\documentclass[aps,pra,twocolumn,superscriptaddress]{revtex4-2}

\usepackage{amsmath}
\usepackage{amssymb}
\usepackage{dcolumn}
\usepackage[english]{babel}
\usepackage{natbib}
\usepackage{graphicx}
\usepackage{xcolor}
\usepackage{braket}

\newcommand{\veps}{\mathcal{E}}
\newcommand{\abs}[1]{\left| #1 \right|}
\newcommand{\dd}{\mathrm{d}}
\newcommand{\ee}{\mathrm{e}}
\newcommand{\ii}{\mathrm{i}}

\begin{document}

\title{Borromean states in a one-dimensional three-body system}

\author{Tobias Schnurrenberger}
\affiliation{German Aerospace Center (DLR), Institute of Quantum Technologies, 89081 Ulm, Germany}

\author{Lucas Happ}
\affiliation{Few-body Systems in Physics Laboratory, RIKEN Nishina Center, Wakō 351-0198, Japan}

\author{Maxim A. Efremov}
\affiliation{German Aerospace Center (DLR), Institute of Quantum Technologies, 89081 Ulm, Germany}

\date{\today}

\begin{abstract}
We show the existence of Borromean bound states in a one-dimensional quantum three-body system composed of two identical bosons and a distinguishable particle. It is assumed that there is no interaction between the two bosons, while the mass-imbalanced two-body subsystems can be tuned to be either bound or unbound. Within the framework of the Faddeev equations, the three-body spectrum and the corresponding wave functions are computed numerically. In addition, we identify the parameter-space region for the two-body interaction, where the Borromean states occur, evaluate their geometric properties, and investigate their dependence on the mass ratio.
\end{abstract}

\maketitle

\section{Introduction}\label{sec:I}
Few-body physics plays a central role in the fields of ultra-cold quantum gases \cite{Blume2012}, nuclear physics \cite{Nielsen2001} and hadron physics \cite{Richard1992}. Special attention is given to three-body systems that have resonantly interacting two-body subsystems because they often show universal properties that are independent of the details of their short range potentials \cite{Braaten2006}. Three-body bound states can even exist in situations when all of the two-body subsystems are unbound and those states are then named \textit{Borromean} states \cite{Zhukov1993,Naidon2017}.

In three spatial dimensions, a two-body system with an overall attractive potential may be unbound, when the interaction is weak enough. A system of three identical bosons with the same pairwise interaction can however be bound. Therefore there is a window for the coupling constant where Borromean states can occur \cite{Goy1995}. The Efimov effect is a well known example to show this behaviour \cite{Efimov1970,Efimov1973} and the theoretical predictions have been verified experimentally \cite{Naidon2017}. Moreover, such Borromean binding plays an essential role in subatomic physics, e.g. in halo nuclei \cite{Zhukov1993,Riisager2013}.

When the particles are restricted to a two dimensional plane, the situation is different. Here a two-body interaction with an overall attractive contribution always supports a bound state \cite{Simon1976}. Therefore the bound states of a three-body system with such two-body interactions are not Borromean. Nevertheless it has been shown that Borromean three-body states can exist in two dimensions by adding a repulsive contribution to the two-body interaction potential \cite{Volosniev2013,Volosniev2014}.

Similar to two dimensions, a one-dimensional two-body system with an overall attractive potential always has a bound state \cite{Simon1976}. This makes it difficult to find situations where Borromean states can occur. To the best of our knowledge, Borromean states have not yet been predicted or observed in a one-dimensional setup. In this paper we consider a two-body interaction that has tunable attractive and repulsive contributions, which are separated in space. In this way, the parameters for this potential can be chosen such that the two-body system is either bound or unbound. Further, our three-body system consist of two types of particles with different masses. We solve this three-body system numerically with the Faddeev equations in momentum space \cite{Happ2022} and find that this three-body system indeed has Borromean states. Moreover, we identify their region of existence in the parameter-space of the two-body interaction and analyze their geometric properties. In addition, we find that the number of Borromean states increases for larger mass ratios.

Considering the experimental feasibility, quasi one-dimensional systems in the form of cigar shaped traps for ultra cold quantum gases have been realized \cite{Bloch2008}. In addition, the interaction between different types of atoms can be tuned with Feshbach resonances \cite{Timmermans1999,Chin2010} or confinement-induced resonances \cite{Dunjko2011,Green2017}. We have performed our calculations for the mass ratio corresponding to a Cs-Li mixture \cite{Pires2014,Tung2014}. By identifying the rate of the one-dimensional three-body recombination processes \cite{Mehta_3b_1d}, we think it will be possible to examine the predicted Borromean states in future experimental studies.

The article is organized as follows. In section \ref{sec:II} we introduce the underlying two-body subsystems as well as the corresponding three-body system. We also discuss the Faddeev equations that we use in our numerical calculations. In section \ref{sec:III} we show from our numerical results the existence of Borromean states and their dependence on the system parameters. In section \ref{sec:IV} we summarize our findings. The Appendices \ref{sec:appxA}, \ref{sec:appxB} and \ref{sec:appxC} provide further details on our calculations.
\section{Two- and three-particle systems}\label{sec:II}

\subsection{Two-particle subsystem}

\begin{figure}[t]
    \centering
    \includegraphics{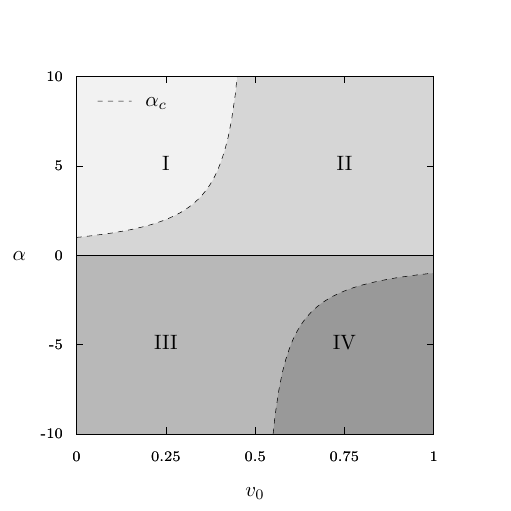}
    \caption{
        The parameter-space $(\alpha, v_0)$ of the two-body subsystems divided into four separate regions with respect to the number of bound and virtual states. From top-left to bottom-right: one virtual state (region I), one bound state (region II), one virtual and one bound state (region III), two bound states (region IV). The dashed line indicates $\alpha_{c}(v_0)$ given by Eq.~\eqref{eqn:alpha_c}. 
    }
	\label{fig:paramspace}
\end{figure}

We start from a one-dimensional two-particle system consisting of distinguishable particles with masses $M$ and $m$. The stationary Schr\"odinger equation for the two-particle wave function $\psi^{(2)}(x)$ is then given by
\begin{align}\label{eqn:2b_schroedinger}
    \left[ -\frac{1}{2}\frac{\dd^2}{\dd x^2} + v(x) \right] \psi^{(2)}(x)=\veps^{(2)}\psi^{(2)}(x),
\end{align}
with an interaction potential $v(x)$. A two-body potential in one-dimension with $\int \dd x ~v(x) < 0$ always has a bound state \cite{Simon1976}. In contrast, a single repulsive barrier does not support a bound state. In order to be able to tune the system between a bound and unbound regime, we need both, an attractive and repulsive term. For simplicity, and in order to treat the two-body system analytically, we consider a potential consisting of two delta distributions
\begin{align}
 \label{eqn:2BPotentialMain}
    v(x)=-v_0\left[ \delta\left(x-\frac{1}{2}\right) -\alpha \delta\left(x+\frac{1}{2}\right) \right]
\end{align}
with the parameters $v_0>0$ and $\alpha$, describing the magnitudes of the overall potential and its relative repulsive contribution respectively.

Both equations~\eqref{eqn:2b_schroedinger} and~\eqref{eqn:2BPotentialMain} are presented in dimensionless variables. The position coordinate $x$ is measured in units of the distance $a$ between the two delta functions. The potential strength $v_0$ and the two-particle energy $\veps^{(2)}$ are both given in units of a characteristic energy $\hbar^2/\mu a^2$, where $\mu=Mm/(M+m)$ is the reduced mass. 

In Appendix \ref{sec:appxA} we solve Eq.~\eqref{eqn:2b_schroedinger} analytically, derive a transcendental equation for the two-particle energy $\veps^{(2)}$ and discuss the number of solutions depending on $\alpha$. For $v_0>0$ we can distinguish four regions in the parameter-space $(\alpha,v_0)$, Fig. \ref{fig:paramspace}, corresponding to different numbers of bound and virtual states \cite{rogernewton} in our two-body subsystem. For $\alpha>0$ the line
\begin{align}\label{eqn:alpha_c}
    \alpha_{c}(v_0)=\frac{1}{1-2v_0}
\end{align}
corresponds to $\veps^{(2)}=0$ and separates the region I with a single virtual state and region II with a single bound state. To explore the Borromean states in the three-particle system, we focus on the transition between these two regions. For $\alpha<0$, the two-particle subsystem supports either one virtual and one bound state (region III), or two bound states (region IV).

\subsection{Three-particle system}

\begin{figure}[t]
    \centering
    \includegraphics{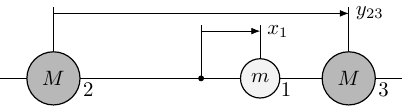}
    \caption{
        The Jacobi coordinates $x_1$ and $y_{23}$ for the one-dimensional three-particle (BBX) system. The dot shows the center-of-mass of two identical bosons, particles $2$ and $3$.
    }
	\label{fig:jacobi} 
\end{figure}

Next, we consider a three-body system in one dimension with two identical bosons (B) of mass $M$ and one distinguishable particle (X) of mass $m$. We assume no interaction between the two bosons, while the BX subsystems interact via the potential $v(x)$, Eq.~\eqref{eqn:2BPotentialMain}.

By eliminating the center-of-mass motion of the three-body BBX system, depicted in Fig. \ref{fig:jacobi}, we arrive at the stationary Schr\"odinger equation
\begin{align}\label{eqn:3bodyschroedinger}
    \left[H_0
        + V_{21}
        + V_{31}
    \right]
    \psi^{(3)} =\veps^{(3)} \psi^{(3)}
\end{align}
for the three-particle wave function $\psi^{(3)}(x_1,y_{23})$, where the Hamiltonian of the free motion is given by
\begin{align}\label{eqn:3bodyH_0}
H_0 =-\frac{\alpha_x}{2}\frac{\dd^2}{\dd x^2_1}
     -\frac{\alpha_y}{2}\frac{\dd^2}{\dd y_{23}^2}
\end{align}
and both of its coefficients
\begin{align}
    \alpha_{x} = \frac{m+2M}{2(m+M)}\quad\text{and}\quad&
    \alpha_{y} = \frac{2m}{m+M}
\end{align}
depend only on the mass ratio $M/m$. The two-body potential terms 
\begin{align}
    V_{21} = v\left(x_1+\frac{y_{23}}{2}\right)\quad\text{and}\quad&
    V_{31} = v\left(x_1-\frac{y_{23}}{2}\right)
\end{align}
for the BX subsystems are given by the function $v(x)$, Eq.~\eqref{eqn:2BPotentialMain}, with the arguments being the relative distances between the respective particles B and X.

In this article, we are interested in finding the Borromean states and their main features in our system. A three-body state is called \textit{Borromean}, if it is bound while all of its two-body subsystems are unbound by themselves \cite{Zhukov1993, Naidon2017}.
For our interaction potential $v(x)$, Eq.~\eqref{eqn:2BPotentialMain}, the Borromean states should therefore lie in the region I of the parameter-space $(\alpha, v_0)$, displayed in Fig. \ref{fig:paramspace}, where there is no two-body bound state.

\subsection{Faddeev equations}
We employ the Faddeev equations \cite{faddeev_scattering_1961, sitenko, Ball1968} in order to calculate the bound-state spectrum and the wave functions of the BBX system. For a derivation of the explicit form of the Faddeev equations for our type of system, we refer to  Appendix A of the paper \cite{Happ2022}. In practice, all information is encapsulated in the set of one-dimensional integral equations
\begin{align}
 \label{eqn:IntegralEqn}
    \varphi_{\lambda}(p,\veps)=~&\sum_{\nu} \int_\mathbb{R} \frac{\dd q}{2\pi} K_{\lambda\nu}(p,q,\veps) \varphi_{\nu}(q,\veps)
\end{align}
for the functions $\varphi_{\nu}(p,\veps)$ with the kernel
\begin{align}
    K_{\lambda\nu}(p,q,\veps)=~&\frac{g_{\lambda}(q+\beta p,\veps_p) \, g^{\ast}_{\nu}(p+\beta q,\veps_q)}{\veps-\frac{1}{2}q^2-\frac{1}{2}p^2-\beta pq} \tau_{\nu}(\veps_q)
\end{align}
and the shorthand notation
\begin{align}
    \beta = \frac{M}{M+m} \quad\mathrm{and}\quad \veps_p = \veps-\frac{1}{2}\alpha_x\alpha_y p^2.
\end{align}

The functions $g_\nu(k,\veps)$ and $\tau_\nu(\veps)$ originate from the separable expansion for the off-shell $t$-matrix of the BX subsystem, whereas the indices $\lambda$ and $\nu$ denote the number of terms in this expansion. As shown in Appendix \ref{sec:appxB}, for the specific form of our interaction potential $v(x)$, Eq.~\eqref{eqn:2BPotentialMain}, the separable expansion has exactly two terms, $\nu \in \{-,+\}$, and $g_\nu(k,\veps)$ and $\tau_\nu(\veps)$ can be derived analytically. 

We solve Eq.~\eqref{eqn:IntegralEqn} numerically, Appendix \ref{sec:appxC}, and obtain the energy $\veps$ of three-body bound states together with their functions $\varphi_\nu(p,\veps)$. The latter can be used to construct the Faddeev component
\begin{equation}
 \label{eqn:faddeevcomp}
    \Phi(k,p) = \sum_{\nu} g_\nu(k,\veps_p) \tau_\nu(\veps_p) \varphi_\nu(p,\veps)
\end{equation}
and finally the wave function
\begin{align}
\label{eqn:wavefunction}
    \psi(p_1,k_{23}) &=~G_0(p_1, k_{23},\veps)\nonumber\\
    &\times\biggl[ \Phi\left(-\alpha_x p_1-\frac{\alpha_y}{2}k_{23},-\frac{1}{2}p_1+k_{23}\right)\nonumber\\
    &+\hspace{1.85mm}\Phi\left(-\alpha_x p_1+\frac{\alpha_y}{2}k_{23},-\frac{1}{2}p_1-k_{23}\right) \biggl]
\end{align}
of the three-body system in momentum space. Here,
\begin{equation}
  G_0(p_1, k_{23},\veps)=\frac{(2\pi)^2}{\veps-\frac{1}{2}\alpha_x p_1^2-\frac{1}{2}\alpha_y k_{23}^2}
\end{equation}
is the Green function of the free three-body system, corresponding to the Hamiltonian $H_0$, Eq.~\eqref{eqn:3bodyH_0}. 

In this way, the wave function in position space can be obtained from the Fourier transform
\begin{equation}
  \psi^{(3)}(x_1,y_{23})=\int_{\mathbb R}\frac{\dd p_1}{2\pi}\int_{\mathbb R}\frac{\dd k_{23}}{2\pi}
                         \ee^{\ii(p_1x_1+k_{23}y_{23})}\psi^{(3)}(p_1,k_{23})
\end{equation}
of the wave function in momentum space.
\section{Borromean states} \label{sec:III}

In this section we study the three-body spectrum of the BBX system and the conditions for the parameters $\alpha$ and $v_0$ under which the  Borromean states exist, as well as the ground state wave function and its geometric properties. Further, we examine the dependence of Borromean states on the mass ratio $M/m$.

The goal is to observe the behavior of the three-body BBX system, while its BX subsystems undergo a transition from supporting exactly one bound state to supporting only a single virtual state. To do this, we are tuning the parameters $\alpha$ and $v_0$ for both two-body interactions $V_{21}$ and $V_{31}$ simultaneously, so that both BX subsystems are identical at all times.

\subsection{Existence and Borromean window}

\begin{figure*}[t]
    \centering
    \includegraphics{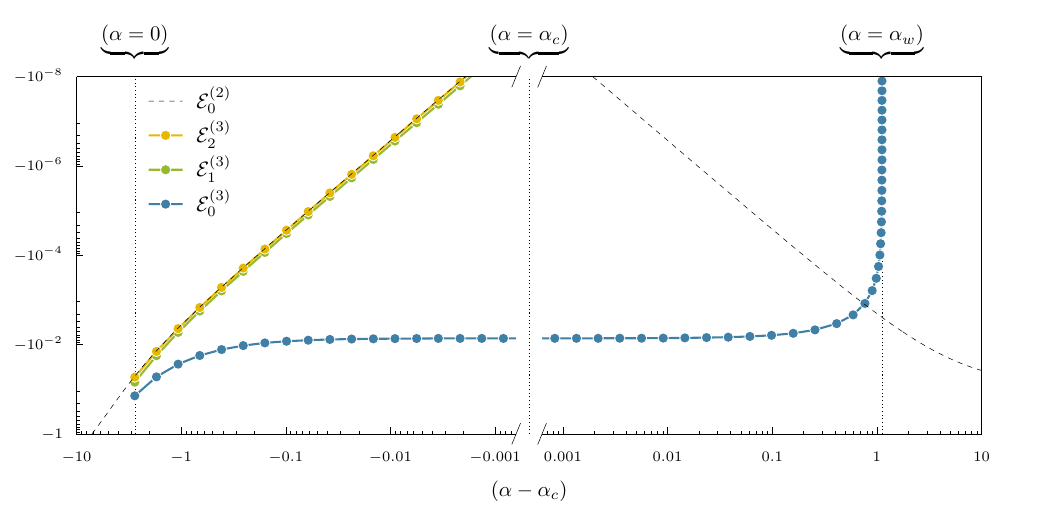}
    \caption{
        Three-body spectrum as a function of the repulsion parameter $\alpha$ for the coupling constant $v_0=0.32$ and the mass ratio $M/m=22.2$. The dashed line shows the energy $\veps^{(2)}_0$ of the two-body bound state (left) or virtual state (right), whereas the dotted lines display the energies $\veps^{(3)}_n$ of the three-body bound states. As $\alpha$ approaches $\alpha_c$ from below (left), the energies $\veps^{(3)}_1$ and $\veps^{(3)}_2$ of the excited three-body states follow the two-body energy described by the power law, Eq.~\eqref{eqn:eps2asymp}, and vanish at $\alpha=\alpha_c$. On the contrary, the three-body ground state has a finite, negative energy $\veps^{(3)}_0$ at $\alpha=\alpha_c$ and remains bound even for $\alpha>\alpha_c$ (right), until $\alpha=\alpha_w$. As for $\alpha > \alpha_c$ there is no two-body bound state, this three-body state is Borromean, and the interval $\alpha_c < \alpha < \alpha_w$ defines the Borromean window.
     }
	\label{fig:spectrum}
\end{figure*}

\begin{figure*}[t]
    \centering
    \includegraphics{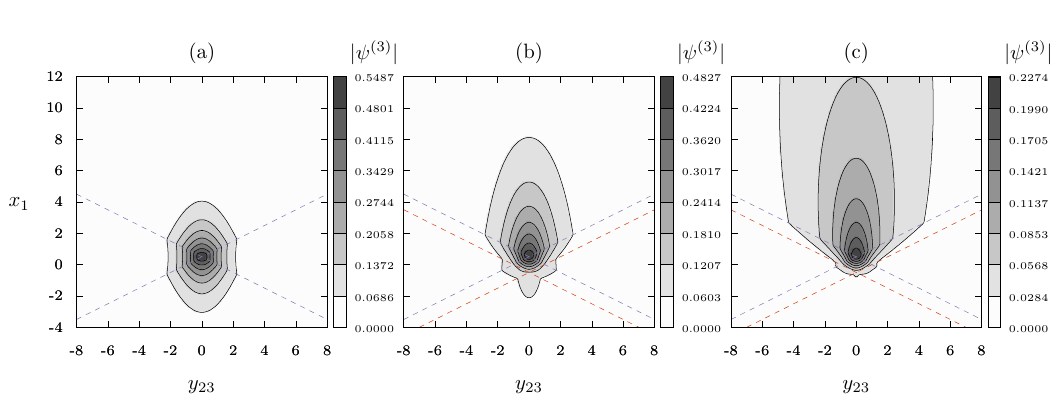}
    \caption{
        Contour plots of the Borromean ground state wave function $|\psi^{(3)}(x_1,y_{23})|$ in position space spanned by the Jacobi coordinates $x_1$ and $y_{23}$, Fig. \ref{fig:jacobi}, for three different values of the repulsion parameter: $\alpha=0$ (a), $\alpha=2.11$ (b), $\alpha=3.84$ (c), as well as $v_0=0.32$ and $M/m=22.2$. With increasing $\alpha$, the wave function starts to become increasingly skewed towards the positive $x_1$ direction, whereas it remains symmetric in $y_{23}$ direction. Similarly to the energy spectrum shown in Fig. \ref{fig:spectrum}, this trend is continuous and no sudden change occurs when the state becomes Borromean. For all values of $\alpha$, the derivative of the wave function jumps at the positions of the attractive (repulsive) contact interactions, highlighted by purple (red) dashed lines.
    }
	\label{fig:wf3}
\end{figure*}

\begin{figure}[t]
    \centering
    \includegraphics{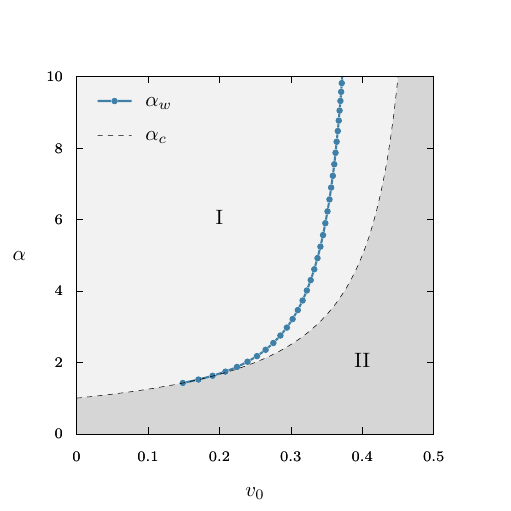}
    \caption{
        Upper-left quadrant of Fig \ref{fig:paramspace}, displaying the parameter space $(\alpha,v_0)$. We highlight two curves: (i) the two-body threshold $\alpha_c$, given by Eq.~\eqref{eqn:alpha_c}, separates regions (I) and (II) with one virtual and one bound state in the BX subsystems, respectively. (ii) The three-body threshold $\alpha_w$ marks the line, above which there is no three-body bound state. It generally depends on the mass ratio, here $M/m = 22.2$, and is determined numerically. Together, both lines define an area (``Borromean window'') between the dotted and dashed curves in which our system has Borromean states.
    }
	\label{fig:bwindow}
\end{figure}

\begin{figure}[t]
    \centering
    \includegraphics{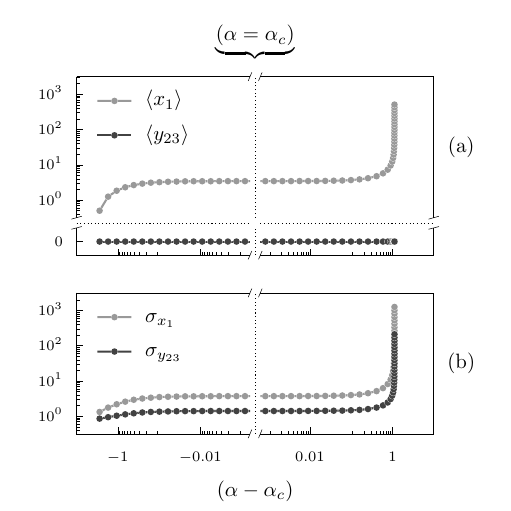}
    \caption{
       The expectation values $\braket{x_1}$, $\braket{y_{23}}$ (a) and the standard deviations $\sigma_{x_1}$, $\sigma_{y_{23}}$ (b), Eq.~\eqref{eqn:standarddeviation}, as functions of $\alpha$. 
       Top: With increasing $\alpha$, also $\braket{x_1}$ increases, meaning that the distinguishable particle is found on average further away from the center-of-mass of the two bosons. Close to the dissociation point $\alpha_w$, $\braket{x_1}$ becomes very large. As the other two particles are identical, $\braket{y_{23}}$ remains zero for all values of $\alpha$. Bottom: both standard deviations $\sigma_{x_1}$, $\sigma_{y_{23}}$ grow with increasing $\alpha$. Overall, the effect is stronger for the $x_1$ coordinate and for both directions strongly enhanced close to the dissociation point $\alpha_w$ where the three-body state becomes very dilute.
    }
	\label{fig:expvalues}
\end{figure}

We start in parameter-region II, Fig. \ref{fig:paramspace}, at the point $(v_0=0.32, \alpha=0)$, where the potential $v(x)$, Eq.~\eqref{eqn:2BPotentialMain}, has only a single delta well and therefore the BX subsystem supports exactly one bound state. We confirm the correctness of our calculations by setting the mass ratio to $M/m=22.2$, corresponding to a Cs-Li mixture, and reproducing the energy spectrum
\begin{align}
        \veps^{(3)}_0=& \enspace 2.7515 \; \veps^{(2)} \nonumber\\
        \veps^{(3)}_1=& \enspace 1.3604 \; \veps^{(2)} \nonumber\\
        \veps^{(3)}_2=& \enspace 1.0525 \; \veps^{(2)} \nonumber
\end{align}
of three-body bound states reported previously in Ref. \cite{Happ2019} and obtained with the Skorniakov-Ter-Martirosian method \cite{stm}. Here, $\veps^{(3)}_0$, $\veps^{(3)}_1$, and $\veps^{(3)}_2$ denote the energies of the ground, first excited, and second excited three-body state, respectively, in units of the two-body energy $\veps^{(2)}$.

Positive values for $\alpha$ introduce a repulsive barrier in the potential $v(x)$, Eq.~\eqref{eqn:2BPotentialMain}. Therefore, increasing $\alpha$ pushes the two-body energy $\veps^{(2)}$ of the single bound state in the BX subsystems closer to the threshold, $\veps^{(2)} = 0$, which is reached at $\alpha=\alpha_c$. For the corresponding BBX system, increasing the barrier strength $\alpha$ results in larger energies $\veps^{(3)}_n$, i.e. weaker binding, for all three bound states, as displayed in Fig. \ref{fig:spectrum}. When $\alpha$ approaches $\alpha_c$, only the energies $\veps^{(3)}_1$ and $\veps^{(3)}_2$ of the excited states vanish together with the energy $\veps^{(2)}$, depicted by the dashed line. The linear behavior in the log-log scale of Fig. \ref{fig:spectrum} suggests a power law dependence, and indeed in Appendix \ref{sec:appxA} we find
\begin{equation}\label{eqn:eps2asymp}
    \veps^{(2)} \approx -\frac12 \left[ \frac{v_0(1-2v_0)^2}{(1-v_0)^2+v_0^2} \right]^2 (\alpha-\alpha_c)^2
\end{equation}
as $\alpha\rightarrow \alpha_c$. In sharp contrast, the three-body ground state does not dissociate. Instead its energy $\veps^{(3)}_0$ takes a finite, negative value at $\alpha=\alpha_c$, Fig. \ref{fig:spectrum}.

By increasing $\alpha$ beyond $\alpha_c$, we enter region I of Fig. \ref{fig:paramspace}, where the two-body systems become unbound and each only has one virtual state. Here, the BBX system is however able to retain the bound ground state. We emphasize that both BX subsystems are unbound, hence the three-body state is Borromean. Moreover, we note that the energy $\veps^{(3)}_0$ of the Borromean state as a function of $\alpha$ changes continuously at the point $\alpha=\alpha_c$. In addition, we observe from Fig. \ref{fig:spectrum}, that for even larger values of $\alpha$, the Borromean state eventually disappears at $\alpha=\alpha_w$, above which there is no three-body bound state anymore. For $\alpha\rightarrow \alpha_w$, we numerically find that $\veps^{(3)}_0$ follows the power law
\begin{equation}
    \veps^{(3)}_0 \approx -0.0014471\;(\alpha_w - \alpha)^{ 1.0417}
\end{equation}
with $\alpha_w = 3.8951$. Here the energy $\veps^{(3)}_0$ is no longer smaller than the energy $\veps^{(2)}$ of the corresponding virtual state in the two-body subsystem.

In summary, we find that for a given potential strength $v_0$ and mass ratio $M/m$, a Borromean state exists in the BBX system in the window $\alpha_c < \alpha < \alpha_w$. By determining $\alpha_c$ and $\alpha_w$ for multiple values of $v_0$, we identify an area in the parameter space $(\alpha, v_0)$, Fig. \ref{fig:paramspace}, where the Borromean state occurs, as depicted in Fig. \ref{fig:bwindow}. Here it becomes evident that a larger magnitude $v_0$ leads to a wider ``Borromean window'' $\left(\alpha_c,\alpha_w\right)$ for $\alpha$.

\subsection{Geometric properties}

Studying the geometric properties of the three-body ground state gives us insight into its particle configuration. For that we calculate the wave function $\psi^{(3)}(x_1,y_{23})$ of the BBX system and present a contour plot of $|\psi^{(3)}(x_1,y_{23})|$ in Fig. \ref{fig:wf3} for three different values of the parameter $\alpha$: $\alpha=0$, (a), $\alpha=2.11$, (b), and $\alpha=3.84$, (c). To visualize the interaction potentials in both BX subsystems, we plot additionally the dashed lines $x_1\pm y_{23}/2=1/2$ (purple) and $x_1\pm y_{23}/2=-1/2$ (red) describing the attractive and repulsive delta wells in $v(x)$, Eq.~\eqref{eqn:2BPotentialMain}, respectively. In addition, we calculate the expectation values $\braket{x_1}$ and $\braket{y_{23}}$, as well as the standard deviations $\sigma_{x_1}$ and $\sigma_{y_{23}}$ as functions of $\alpha$, with
\begin{equation} \label{eqn:standarddeviation}
    \sigma_{z} = \sqrt{\braket{z^2}-\braket{z}^2},
\end{equation}
and display them in Figs. \ref{fig:expvalues} (a) and (b), accordingly.

For zero repulsion in the interaction ($\alpha=0$), Fig. \ref{fig:wf3} (a) shows that the wave function $|\psi^{(3)}(x_1,y_{23})|$ is point symmetric with respect to its expectation values $\braket{x_1}=1/2$ and $\braket{y_{23}}=0$. Since $y_{23}$ is the distance between the identical bosonic particles, Fig. \ref{fig:jacobi}, the wave function $\psi^{(3)}(x_1,y_{23})$ is symmetric in this coordinate, i.e. $\psi^{(3)}(x_1,-y_{23})=\psi^{(3)}(x_1,y_{23})$, giving rise to $\braket{y_{23}}=0$ for all parameters. Apart from the shift by 1/2 in $x_1$ direction, this result coincides with the one found previously \cite{Happ2019} for the same three-body system but with the single contact interaction being centered.

Next, we increase the repulsion, by setting $\alpha=2.11$, a bit below $\alpha_c=2.78$, and present the resulting wave function in Fig. \ref{fig:wf3} (b). Here we have a shift of the expected position $\braket{x_1}=2.31$ of the distinguishable particle, whereas $\braket{y_{23}}=0$, as shown in Fig. \ref{fig:expvalues} (a). Moreover, the wave function becomes broader compared to the case with $\alpha=0$, Fig. \ref{fig:wf3} (a), which is also indicated by the scale of $|\psi(x_1,y_{23})|$ as well as the standard deviations $\sigma_{x_1}$ and $\sigma_{y_{23}}$, presented in Fig. \ref{fig:expvalues} (b). The result also agrees with the fact that the ground state energy $\veps^{(3)}_0$ at $\alpha=2.11$ is smaller by about one order in comparison with $\veps^{(3)}_0$ at $\alpha=0$, as displayed by Fig. \ref{fig:spectrum}.

Now increasing $\alpha$ to the value slightly below $\alpha_w$, which identifies the Borromean three-body threshold, Fig. \ref{fig:wf3} (c), the wave function $\psi(x_1,y_{23})$ shows that the distinguishable particle  moves on average even further away, $\braket{x_1}=19.5$, from the center-of-mass of the identical particles, whereas $\braket{y_{23}}=0$, Fig. \ref{fig:expvalues} (a). In addition, the wave function becomes again broader and extremely delocalized with very weak binding energy $\veps^{(3)}_0$ as $\alpha\rightarrow \alpha_w$, Fig. \ref{fig:spectrum}, before the Borromean ground state dissociates completely. This observation is in line with both standard deviations $\sigma_{x_1}$ and $\sigma_{y_{23}}$ becoming very large when $\alpha$ approaches $\alpha_w$, Fig. \ref{fig:expvalues} (b).

It is important to note, that the wave function continuously changes as the function of $\alpha$ throughout Figs. \ref{fig:wf3} (a), (b) and (c), as do the expectation values presented in Fig. \ref{fig:expvalues}.

\subsection{Dependence on the mass ratio}

\begin{figure}[t]
    \centering
    \includegraphics{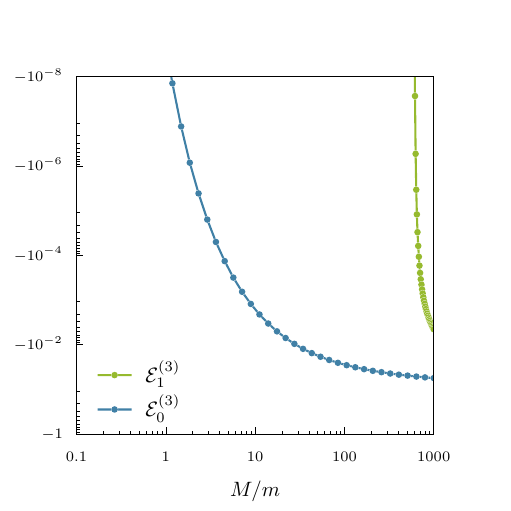}
    \caption{
         Three-body spectrum as a function of the mass ratio $M/m$ and with the coupling constant $v_0=0.32$. We choose the repulsion parameter $\alpha$ just above $\alpha_c$, i.e. inside the Borromean window, Fig. \ref{fig:bwindow}, and therefore the displayed states are Borromean. With increasing mass ratio, the system goes from supporting zero, to one and to two Borromean bound states.
    }
	\label{fig:threshold_massr}
\end{figure}

\begin{figure}[t]
    \centering
    \includegraphics{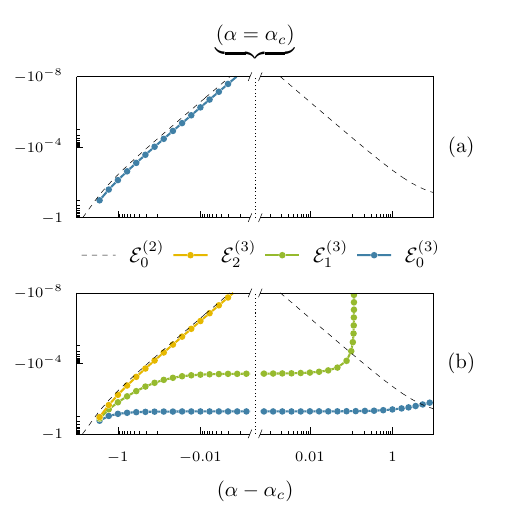}
    \caption{
        Three-body spectrum as a function of $\alpha$ with constant $v_0=0.32$ for the mass ratios $M/m=0.2$ (a) and $M/m=720$ (b). Top: The BBX systems supports only a single three-body bound state for $\alpha<\alpha_c$ and it dissociates at $\alpha=\alpha_c$, leading to no Borromean state for the small mass ratio. Bottom: For the much larger mass ratio, the three-body system supports many bound states for $\alpha < \alpha_c$. Here, only the first three bound states with the lowest energies are presented. The two deepest bound states remain bound also for $\alpha > \alpha_c$ and are hence Borromean states.
    }
	\label{fig:spectrum_2ndb}
\end{figure}

Finally, we analyze the dependence of the Borromean three-body energy on the mass ratio $M/m$. This is shown in Fig. \ref{fig:threshold_massr}. Here we choose $\alpha$ to be just slightly above $\alpha_c$, i.e. we are just inside the Borromean window, where the Borromean states are relatively deeply bound. Nevertheless, we see that for smaller mass ratios, the binding energy of the Borromean ground state (blue line) quickly approaches zero. On the other hand, as the mass ratio increases, this Borromean state becomes more strongly bound. Moreover, when the mass ratio is sufficiently large ($M/m \gtrsim 600$), a second (excited) Borromean state appears, as depicted by a green line. We note that the energy $\veps^{(3)}$ is expressed in units of $\hbar^2/(\mu a^2)$ with the reduced mass $\mu$. Overall, for larger mass ratios the three-body states become more bound, and the Borromean states appear one by one.

To trace back the origin of the second Borromean state, we analyze the three-body spectrum for different values of $\alpha$ and $M/m$, as demonstrated in Fig. \ref{fig:spectrum_2ndb}. In subfigure (a) we choose $M/m = 0.2$ and we cannot find a Borromean state. There is just a single non-Borromean bound state for $\alpha<\alpha_c$. That's similar to the case, when the subsystems have a zero-range interaction, then the three-body system does not support more than one bound state for $M/m\leq1$ \cite{Kartavtsev2009}. At an intermediate value of the mass ratio, e.g. $M/m = 22.2$ in Fig. \ref{fig:spectrum}, we clearly see the three-body ground state in the Borromean region, $\alpha_c < \alpha < \alpha_w$. Finally, for $M/m = 720$, Fig. \ref{fig:spectrum_2ndb} (b), the ground state is quite deeply bound and we see the second state (green line) in the Borromean window. Again, the energy of all three-body bound states changes smoothly as a function of $\alpha$ when crossing the value $\alpha=\alpha_c$.
\section{Conclusion}\label{sec:IV}
In this article we have investigated Borromean states in a three-body BBX system confined in one dimension, provided no interaction in the BB subsystem and each BX subsystem supports only a single virtual state. By solving the Faddeev equations numerically, we have calculated the spectrum and the corresponding wave functions of the BBX system. We have demonstrated for the first time the existence of Borromean states in 1D and, for a given mass ratio, identified an area in the parameter-space, where these states occur. Further, we have found that the number of Borromean states and their binding energies strongly depend on the mass ratio of the two particles types. In addition, we have shown that in all cases these novel states originate from ordinary bound ones with the lowest three-body binding energies.

As an outlook, for experimental verification of our theoretical predictions and their applications in physics of quantum gases, it is useful to calculate the rate of the one-dimensional three-body recombination processes for the BBX system considered here in detail. Moreover, further theoretical studies of Borromean states in one dimension might involve the use of more realistic atom-atom potentials, like the Lennard-Jones potential, that has an attractive and repulsive contribution.

\section*{Acknowledgment}
We thank A. Volosniev and P. Belov for fruitful discussions and advice. The work of L.~H. is gratefully supported by the RIKEN special postdoctoral researcher program. The authors gratefully acknowledge the scientific support and HPC resources provided by the German Aerospace Center (DLR). The HPC system CARA is partially funded by “Saxon State Ministry for Economic Affairs, Labour and Transport“ and „Federal Ministry for Economic Affairs and Climate Action”.

\begin{appendix}
    \section{BX subsystem}\label{sec:appxA}

\begin{figure}[t]
    \centering
    \includegraphics{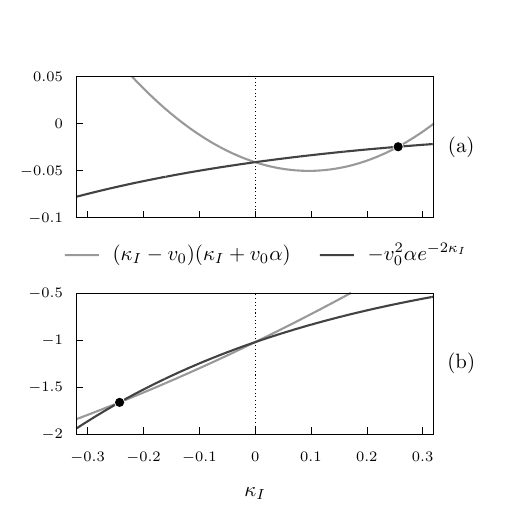}
    \caption{
        The left- and right-hand sides of the transcendental equation~\eqref{eqn:2body_teqn} as  functions of $\kappa_I$ with $v_0=0.32$. We look for solutions $\kappa_I\neq0$. In subfigure (a) $\alpha=0.4$ and the intersection  of these lines is marked by a dot and occurs for $\kappa_I>0$, corresponding to the bound state. In subfigure (b) $\alpha=10$, the intersection occurs for $\kappa_I<0$ and corresponds to the virtual state.
    }
	\label{fig:teqn}
\end{figure}

\begin{figure}[t]
    \centering
    \includegraphics{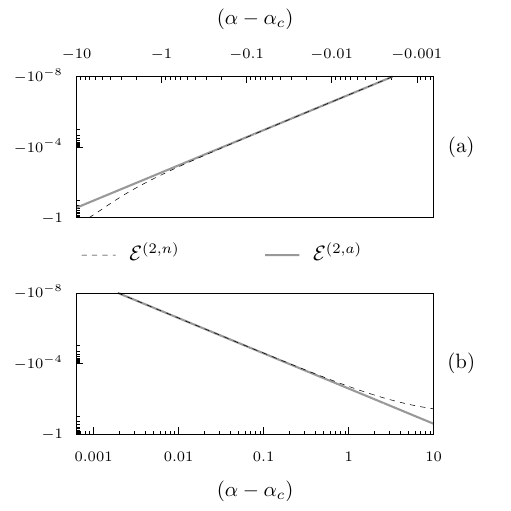}
    \caption{
        Comparison of the two-body energy $\veps^{(2,a)}$, given by the asymptotic expansion, Eq.~\eqref{eqn:eps2asymp_c1}, to $\veps^{(2,n)}$ obtained as the numerical solution of the transcendental equation~\eqref{eqn:2body_teqn}. Subfigure (a) shows the solution for the bound state with $\alpha<\alpha_c$, whereas subfigure (b) displays the solution for the virtual state with $\alpha_c<\alpha$.
    }
	\label{fig:eps2}
\end{figure}

In this Appendix we derive the transcendental equation for the spectrum of the BX subsystem. Moreover, we discuss the number of solutions as a function of $\alpha$ and derive the asymptotic behavior of the two-body energy near the threshold $\alpha \to \alpha_c$.

We start from the stationary Schr\"odinger equation
\begin{align}\label{eqn:2BSchroedingerAppx}
    \left[ -\frac{1}{2}\frac{\dd^2}{\dd x^2} + v(x) \right] \psi^{(2)}(x)
    = \veps^{(2)} \psi^{(2)}(x),
\end{align}
where the two-body interaction potential $v(x)$ is given by the potential
\begin{align}\label{eqn:2BPotentialAppx}
    v(x)=-v_0\left[
         \delta\left(x-\frac{1}{2}\right)
         -\alpha\delta\left(x+\frac{1}{2}\right)
    \right]
\end{align}
with positive $v_0$.

With $\kappa$ defined as $\veps^{(2)}=\kappa^2/2$, the solution of Eq.~\eqref{eqn:2BSchroedingerAppx} can be written in the form
\begin{align}
    \psi^{(2)}(x)=
    \begin{cases}
        A\ee^{\ii\kappa x}+A^{\prime}\ee^{-\ii\kappa x}, & x<-\frac{1}{2}\\
        B\ee^{\ii\kappa x}+B^{\prime}\ee^{-\ii\kappa x}, & -\frac{1}{2}<x<\frac{1}{2}\\
        C\ee^{\ii\kappa x}+C^{\prime}\ee^{-\ii\kappa x}, & \frac{1}{2}<x,
    \end{cases}
\end{align}
where $A$, $A^{\prime}$, $B$, $B^{\prime}$, $C$, and $C^{\prime}$ are constants.

To determine them, we first incorporate the outgoing-wave boundary condition \cite{RayaZavin_2004}, that is $A=C^{\prime}=0$. Next, we require the wave function to be continuous at $x=-\frac{1}{2}$ and $x=\frac{1}{2}$ and arrive at two conditions
\begin{align}
    A^{\prime}\ee^{\ii\kappa/2} =& \; B\ee^{-\ii\kappa/2} + B^{\prime}\ee^{\ii\kappa/2}\label{eqn:2bodycoeff1}\\
    B\ee^{\ii\kappa/2} + B^{\prime}\ee^{-\ii\kappa/2} =& \; C\ee^{\ii\kappa/2}.
\end{align}

Integrating the Schr\"odinger equation~\eqref{eqn:2BSchroedingerAppx} on the intervals $x\in\left[-\frac12-\epsilon,-\frac12+\epsilon\right]$ and $x\in\left[\frac12-\epsilon,\frac12+\epsilon\right]$, and then taking the limit $\epsilon \to 0$, gives us expressions for the jump of the derivative of the wave function at $x=-\frac12$ and $x=\frac12$, respectively. These expressions yield the two additional conditions
\begin{align}
    A^{\prime}\left(1+2\ii\frac{v_0\alpha}{\kappa}\right)\ee^{\ii\kappa/2} + B\ee^{-\ii\kappa/2} - B^{\prime}\ee^{\ii\kappa/2} =& \; 0\\
     -B\ee^{\ii\kappa/2} + B^{\prime}\ee^{-\ii\kappa/2} + C\left(1-2\ii\frac{v_0}{\kappa}\right)\ee^{\ii\kappa/2} =& \; 0.\label{eqn:2bodycoeff4}
\end{align}

The system of four linear equations~\eqref{eqn:2bodycoeff1}--\eqref{eqn:2bodycoeff4} for the coefficients $A^{\prime}$, $B$, $B^{\prime}$, and $C$ has non-trivial solutions only if the determinant of this system equals zero, giving rise to a transcendental equation for $\kappa=\kappa_{R}+\ii\kappa_{I}$. By setting $\kappa_{R}=0$ we restrict the possible solutions to virtual ($\kappa_I<0$) and bound ($\kappa_I>0$) states and obtain the transcendental equation
\begin{align}\label{eqn:2body_teqn}
    (\kappa_{I}-v_0)(\kappa_{I}+v_0\alpha)=-v_0^2\alpha \ee^{-2\kappa_{I}}
\end{align}
for $\kappa_I$.

We are not interested in the trivial solution $\kappa_{I}=0$ of Eq.~\eqref{eqn:2body_teqn}. In Fig. \ref{fig:teqn}, both the left (gray line) and right (black line) sides of Eq.~\eqref{eqn:2body_teqn} are depicted as a function of $\kappa_I$. We see that for $\alpha>0$ there is one non-trivial solution (black dot). Whether this root is located on the positive or negative $\kappa_{I}$-axis depends on the difference of the derivatives of the left- and right-hand sides of Eq.~\eqref{eqn:2body_teqn} at $\kappa_I=0$, that is $v_0(\alpha-1)-2v_0^2\alpha=v_0(\alpha-1-2v_0\alpha)$. This difference vanishes for
\begin{align}
    \alpha = \alpha_{c} = \frac{1}{1-2v_0}.
\end{align}
In this way, for $0 < \alpha < \alpha_{c}$ there is only one bound state (with $\kappa_I>0$) in the BX subsystem, whereas for $0 < \alpha_c < \alpha$, the subsystem supports only one virtual state (with $\kappa_I<0$). In the case $\alpha<0$, there are one bound and one virtual state for $\alpha_{c} < \alpha < 0$ and two bound states for $\alpha < \alpha_c < 0$. All these cases are summarized in the parameter space of the BX subsystem, Fig. \ref{fig:paramspace}. 

In addition, we now derive the analytical formula for the non-trivial solution of Eq.~\eqref{eqn:2body_teqn} as $\alpha\rightarrow \alpha_c$. In this limit the non-trivial root $\kappa_I\rightarrow 0$. Therefore, we can use the asymptotic expansion 
\begin{align}
    e^{-2\kappa_I}=1-2\kappa_I+2\kappa_I^2+{\mathcal O}(\kappa_I^3)
\end{align}
in Eq.~\eqref{eqn:2body_teqn} and solve the resulting equation for $\kappa_I\neq 0$, to arrive at
\begin{align}
    \kappa_I(\alpha)\approx v_0 \frac{1+\alpha(2v_0-1)}{1+2v_0^2\alpha}.
\end{align}

Next, we expand $\alpha$ around $\alpha_c$ to the first order and obtain
\begin{align}
     \kappa_I(\alpha)\approx c_1 (\alpha-\alpha_c)
\end{align}
with the coefficient
\begin{align}
    c_1=-\frac{v_0(1-2v_0)^2}{(1-v_0)^2+v_0^2}. 
\end{align}
As a result, we obtain the approximate behavior  
\begin{align}\label{eqn:eps2asymp_c1}
    \veps^{(2)}(\alpha)=-\frac{1}{2}\kappa_I^2\approx- \frac{c_1^2}{2} (\alpha-\alpha_c)^2 
\end{align}
of the two-body energy as $\alpha\rightarrow\alpha_c$.
    \section{Separable expansion}\label{sec:appxB}
In this Appendix we derive the analytical form of the separable expansion \cite{sitenko} for the two-body $t$-matrix, corresponding to the two-body potential $v(x)$, Eq.~\eqref{eqn:2BPotentialMain}. 

We expand the $t$-matrix in separable terms
\begin{align}\label{eqn:tmatrix}
    t(k,k^{\prime},\veps)=\sum_{\nu} \tau_\nu(\veps)
    g^{\ast}_\nu(k,\veps) g_\nu(k^{\prime},\veps)
\end{align}
with
\begin{align}
    \tau_\nu(\veps)=\frac{\eta_\nu(\veps)}{\eta_\nu(\veps)-1},
\end{align}
where the functions $\eta_\nu(k,\veps)$ and $g_\nu(\veps)$ are determined by the integral equation
\begin{align}\label{eqn:g_eta_def}
    \int_{\mathbb{R}} \frac{\dd k^{\prime}}{2\pi}\frac{v(k,k^{\prime})}{\veps-\frac{1}{2}k^{\prime 2}}g_{\nu}(k^{\prime},\veps)=\eta_\nu (\veps) g_\nu(k,\veps).
\end{align}
Here,
\begin{align}\label{eqn:mompotential-definition}
    v(k,k^{\prime})=\int_{\mathbb R} \dd x \, v(x) \, \ee^{-\ii(k-k^\prime)x}
\end{align}
is the momentum representation of the two-body interaction potential $v(x)$ and the functions $g_\nu(k,\veps)$ are orthogonal and normalized according to the condition
\begin{align}\label{eqn:g_orthonormal}
    \int_{\mathbb{R}} \frac{\dd k}{2\pi} \frac{g_\nu (k,\veps) \, g_{\nu^\prime}^\ast (k,\veps)}{\veps-\frac{1}{2}k^2} = -\delta_{\nu,\nu^\prime}.
\end{align}

By inserting the potential defined in Eq.~\eqref{eqn:2BPotentialMain} into Eq.~\eqref{eqn:mompotential-definition}, we find 
\begin{align}
    v(k,k^{\prime}) = -v_0\left[ e^{-\ii(k-k^\prime)/2} - \alpha \, \ee^{\ii(k-k^\prime)/2} \right],
\end{align}
resulting in a separable kernel of the integral equation~\eqref{eqn:g_eta_def}. This allows us to rewrite Eq.~\eqref{eqn:g_eta_def} in the form
\begin{align}\label{eqn:g_eta_defG1G2}
    -v_0\left[ \ee^{-\ii k/2} G_{\nu}^{(+)}(\veps) -\alpha \, \ee^{\ii k/2} G_{\nu}^{(-)}(\veps) \right] = \eta_\nu (\veps) g_\nu (k,\veps),
\end{align}
where
\begin{align}
    G_{\nu}^{(\pm)}(\veps) =\int_\mathbb{R}\frac{\dd k^{\prime}}{2\pi}\frac{\ee^{\pm\ii k^\prime /2}}{\veps-\frac{1}{2}k^{\prime 2}}g_{\nu}(k^{\prime},\veps)\label{eqn:G1}
\end{align}
are single-argument functions of the energy $\veps$.

In order to determine $G_{\nu}^{(\pm)}(\veps)$, we insert the function $g_\nu (k,\veps)$ given by Eq.~\eqref{eqn:g_eta_defG1G2} into Eq.~\eqref{eqn:G1} and obtain
\begin{align}
 \label{eqn:systemG1G2-1}
    \left(1+\frac{v_0}{\eta_{\nu}}{\mathcal A}\right)G_{\nu}^{(+)}-\frac{v_0\alpha}{\eta_{\nu}}{\mathcal B}G_{\nu}^{(-)}=0\\ 
 \label{eqn:systemG1G2-2}
    \frac{v_0}{\eta_{\nu}}{\mathcal B}^{*}G_{\nu}^{(+)}+\left(1-\frac{v_0\alpha}{\eta_{\nu}}{\mathcal A}\right)G_{\nu}^{(-)}=0
\end{align}
with
\begin{align}
 \label{eqn:parameterA}
    {\mathcal A}=&\int_\mathbb{R}\frac{\dd k^{\prime}}{2\pi}\frac{1}{\veps-\frac{1}{2}k^{\prime 2}}=-\frac{1}{\sqrt{-2\veps}}\\
\intertext{and}
 \label{eqn:parameterB}
    {\mathcal B}=&\int_\mathbb{R}\frac{\dd k^{\prime}}{2\pi}\frac{\ee^{\ii k^\prime}}{\veps-\frac{1}{2}k^{\prime 2}}=-\frac{1}{\sqrt{-2\veps}}\ee^{-\sqrt{-2\veps}},
\end{align}
valid for $\veps<0$.

The system of algebraic equations~\eqref{eqn:systemG1G2-1}--\eqref{eqn:systemG1G2-2} for $G_{\nu}^{(-)}$ and $G_{\nu}^{(+)}$ has non-trivial solutions only when the determinant of this system is zero, that is
\begin{equation}\label{eqn:etadef}
    \left(\eta_\nu+v_0{\mathcal A}\right)\left(\eta_{\nu}-v_0 \alpha{\mathcal A}\right)+v_0^2 \alpha |{\mathcal B}|^2=0,
\end{equation}
resulting in two eigenvalues
\begin{align}
    \eta_{\pm}(\veps) =& \frac{v_0}{2\sqrt{-2\veps}} 
    \left[ 1-\alpha \pm \mathcal{S}(\veps)
    \right]
\end{align}
with
\begin{equation}
    \mathcal{S}(\veps) = \sqrt{(1-\alpha)^2+4\alpha
    \left( 1-e^{-2\sqrt{-2\veps}}\right)}
\end{equation}

Finally, using Eq.~\eqref{eqn:g_eta_defG1G2} together with the normalization condition Eq.~\eqref{eqn:g_orthonormal}, we obtain the expression 
\begin{align}
    g_\nu(k,\veps) =&
    \frac
    {
    \frac{v_0}{\eta_\nu}\abs{\frac{\eta_\nu}{v_0}}(-2\veps)^{1/4}
    }{
    \sqrt{ \ee^{2\sqrt{-2\veps}}\mathcal{P}^2_\nu(\veps) - 2\mathcal{P}_\nu(\veps) + 1 }
    }\nonumber\\
    &\times
    \left[\ee^{\sqrt{-2\veps}}\mathcal{P}_\nu(\veps) \ee^{\ii k/2} - \ee^{-\ii k/2} \right]
\end{align}
with
\begin{equation}
    \mathcal{P}_\nu(\veps)
    = \left( 1 - \frac{\eta_\nu}{v_0}\sqrt{-2\veps} \right)
    = \frac
    {
    \alpha \ee^{-2\sqrt{-2\veps}}
    }{
    \left( \frac{ \eta_\nu}{v_0}\sqrt{-2\veps} + \alpha \right)
    }
\end{equation}
for the corresponding eigenfunctions. Where we used Eq.~\eqref{eqn:etadef} in the second step.
    \section{Numerics}\label{sec:appxC}
In this Appendix we discuss our approach for the numerical solution of the integral Eq.~\eqref{eqn:IntegralEqn}. We use discrete momenta $p_i$ and $q_j$ with the indices $i,j\in\{0,\dots,N-1\}$, with $N$ being the number of discretization points, to define the vector elements
\begin{align}
    \left[ \varphi_{\lambda}(\veps) \right]_i =~& \varphi_{\lambda}(p_i,\veps)\\
    \intertext{and the matrix elements}
    \left[ \tilde K_{\lambda\nu}(\veps) \right]_{ij} =~& \frac{w_j}{2\pi} K_{\lambda\nu}(p_i,q_j,\veps),
\end{align}
respectively. The integral weights $w_j$ depend on the exact form of the discretization. We use the Gauss-Legendre quadrature rule \cite{Delves_Mohamed_1985,Golub1969}. The indices $\lambda$ and $\nu$ label the terms from the separable expansion given in Eq.~\eqref{eqn:tmatrix}. In practice, the sum over $\nu$ in Eq.~\eqref{eqn:IntegralEqn} needs to be finite or truncated. In the case of the two-body potential, Eq.~\eqref{eqn:2BPotentialMain}, we have exactly two terms. The integral equation now takes the form of an eigenvalue problem
\begin{equation}\label{eqn:eigenvalueproblem}
    \underbrace{
    \begin{bmatrix}
        \varphi_0\\
        \varphi_1
    \end{bmatrix}
    }_{=\mathbf{v}}
    =\pm
    \underbrace{
    \begin{bmatrix}
        \tilde K_{00} & \tilde K_{01} \\
        \tilde K_{10} & \tilde K_{11}
    \end{bmatrix}
    }_{=\mathbf{W}}
    \begin{bmatrix}
        \varphi_0\\
        \varphi_1
    \end{bmatrix}
\end{equation}
with fixed eigenvalue $+1$ $(-1)$ for bosons (fermions). Equation~\eqref{eqn:eigenvalueproblem} has a solution when
\begin{equation}\label{eqn:determinant}
    \mathrm{det}\left[\pm
    \mathbf{W}(\veps)
    -\mathbb{I} \right] = 0
\end{equation}
is fulfilled. The determinant in Eq.~\eqref{eqn:determinant} is calculated for a range of energy values $\veps$ and each zero point of the determinant corresponds to an eigen-energy $\veps^{(3)}$ of the three-body spectrum of Eq.~\eqref{eqn:3bodyschroedinger}.

To find the wave function, we calculate the eigenvector $\mathbf{v}(\veps^{(3)})$ of the matrix $\mathbf{W}(\veps^{(3)})$ for the corresponding eigen-energy $\veps^{(3)}$. From the subvectors $\varphi_{\nu}(\veps^{(3)})$ of the eigenvector $\mathbf{v}(\veps^{(3)})$ we can get the Faddeev component from its separable terms via Eq.~\eqref{eqn:faddeevcomp}. Finally, we use Eq.~\eqref{eqn:wavefunction} to obtain the wave function in momentum space.
\end{appendix}


\begin{thebibliography}{32}%
\makeatletter
\providecommand \@ifxundefined [1]{%
 \@ifx{#1\undefined}
}%
\providecommand \@ifnum [1]{%
 \ifnum #1\expandafter \@firstoftwo
 \else \expandafter \@secondoftwo
 \fi
}%
\providecommand \@ifx [1]{%
 \ifx #1\expandafter \@firstoftwo
 \else \expandafter \@secondoftwo
 \fi
}%
\providecommand \natexlab [1]{#1}%
\providecommand \enquote  [1]{``#1''}%
\providecommand \bibnamefont  [1]{#1}%
\providecommand \bibfnamefont [1]{#1}%
\providecommand \citenamefont [1]{#1}%
\providecommand \href@noop [0]{\@secondoftwo}%
\providecommand \href [0]{\begingroup \@sanitize@url \@href}%
\providecommand \@href[1]{\@@startlink{#1}\@@href}%
\providecommand \@@href[1]{\endgroup#1\@@endlink}%
\providecommand \@sanitize@url [0]{\catcode `\\12\catcode `\$12\catcode
  `\&12\catcode `\#12\catcode `\^12\catcode `\_12\catcode `\%12\relax}%
\providecommand \@@startlink[1]{}%
\providecommand \@@endlink[0]{}%
\providecommand \url  [0]{\begingroup\@sanitize@url \@url }%
\providecommand \@url [1]{\endgroup\@href {#1}{\urlprefix }}%
\providecommand \urlprefix  [0]{URL }%
\providecommand \Eprint [0]{\href }%
\providecommand \doibase [0]{https://doi.org/}%
\providecommand \selectlanguage [0]{\@gobble}%
\providecommand \bibinfo  [0]{\@secondoftwo}%
\providecommand \bibfield  [0]{\@secondoftwo}%
\providecommand \translation [1]{[#1]}%
\providecommand \BibitemOpen [0]{}%
\providecommand \bibitemStop [0]{}%
\providecommand \bibitemNoStop [0]{.\EOS\space}%
\providecommand \EOS [0]{\spacefactor3000\relax}%
\providecommand \BibitemShut  [1]{\csname bibitem#1\endcsname}%
\let\auto@bib@innerbib\@empty
\bibitem [{\citenamefont {Blume}(2012)}]{Blume2012}%
  \BibitemOpen
  \bibfield  {author} {\bibinfo {author} {\bibfnamefont {D.}~\bibnamefont
  {Blume}},\ }\bibfield  {title} {\bibinfo {title} {Few-body physics with
  ultracold atomic and molecular systems in traps},\ }\href
  {https://doi.org/10.1088/0034-4885/75/4/046401} {\bibfield  {journal}
  {\bibinfo  {journal} {Reports on Progress in Physics}\ }\textbf {\bibinfo
  {volume} {75}},\ \bibinfo {pages} {046401} (\bibinfo {year}
  {2012})}\BibitemShut {NoStop}%
\bibitem [{\citenamefont {Nielsen}\ \emph {et~al.}(2001)\citenamefont
  {Nielsen}, \citenamefont {Fedorov}, \citenamefont {Jensen},\ and\
  \citenamefont {Garrido}}]{Nielsen2001}%
  \BibitemOpen
  \bibfield  {author} {\bibinfo {author} {\bibfnamefont {E.}~\bibnamefont
  {Nielsen}}, \bibinfo {author} {\bibfnamefont {D.}~\bibnamefont {Fedorov}},
  \bibinfo {author} {\bibfnamefont {A.}~\bibnamefont {Jensen}},\ and\ \bibinfo
  {author} {\bibfnamefont {E.}~\bibnamefont {Garrido}},\ }\bibfield  {title}
  {\bibinfo {title} {The three-body problem with short-range interactions},\
  }\href {https://doi.org/https://doi.org/10.1016/S0370-1573(00)00107-1}
  {\bibfield  {journal} {\bibinfo  {journal} {Physics Reports}\ }\textbf
  {\bibinfo {volume} {347}},\ \bibinfo {pages} {373} (\bibinfo {year}
  {2001})}\BibitemShut {NoStop}%
\bibitem [{\citenamefont {Richard}(1992)}]{Richard1992}%
  \BibitemOpen
  \bibfield  {author} {\bibinfo {author} {\bibfnamefont {J.-M.}\ \bibnamefont
  {Richard}},\ }\bibfield  {title} {\bibinfo {title} {The nonrelativistic
  three-body problem for baryons},\ }\href
  {https://doi.org/https://doi.org/10.1016/0370-1573(92)90078-E} {\bibfield
  {journal} {\bibinfo  {journal} {Physics Reports}\ }\textbf {\bibinfo {volume}
  {212}},\ \bibinfo {pages} {1} (\bibinfo {year} {1992})}\BibitemShut {NoStop}%
\bibitem [{\citenamefont {Braaten}\ and\ \citenamefont
  {Hammer}(2006)}]{Braaten2006}%
  \BibitemOpen
  \bibfield  {author} {\bibinfo {author} {\bibfnamefont {E.}~\bibnamefont
  {Braaten}}\ and\ \bibinfo {author} {\bibfnamefont {H.-W.}\ \bibnamefont
  {Hammer}},\ }\bibfield  {title} {\bibinfo {title} {Universality in few-body
  systems with large scattering length},\ }\href
  {https://doi.org/https://doi.org/10.1016/j.physrep.2006.03.001} {\bibfield
  {journal} {\bibinfo  {journal} {Physics Reports}\ }\textbf {\bibinfo {volume}
  {428}},\ \bibinfo {pages} {259} (\bibinfo {year} {2006})}\BibitemShut
  {NoStop}%
\bibitem [{\citenamefont {Zhukov}\ \emph {et~al.}(1993)\citenamefont {Zhukov},
  \citenamefont {Danilin}, \citenamefont {Fedorov}, \citenamefont {Bang},
  \citenamefont {Thompson},\ and\ \citenamefont {Vaagen}}]{Zhukov1993}%
  \BibitemOpen
  \bibfield  {author} {\bibinfo {author} {\bibfnamefont {M.}~\bibnamefont
  {Zhukov}}, \bibinfo {author} {\bibfnamefont {B.}~\bibnamefont {Danilin}},
  \bibinfo {author} {\bibfnamefont {D.}~\bibnamefont {Fedorov}}, \bibinfo
  {author} {\bibfnamefont {J.}~\bibnamefont {Bang}}, \bibinfo {author}
  {\bibfnamefont {I.}~\bibnamefont {Thompson}},\ and\ \bibinfo {author}
  {\bibfnamefont {J.}~\bibnamefont {Vaagen}},\ }\bibfield  {title} {\bibinfo
  {title} {Bound state properties of borromean halo nuclei: 6he and 11li},\
  }\href {https://doi.org/https://doi.org/10.1016/0370-1573(93)90141-Y}
  {\bibfield  {journal} {\bibinfo  {journal} {Physics Reports}\ }\textbf
  {\bibinfo {volume} {231}},\ \bibinfo {pages} {151} (\bibinfo {year}
  {1993})}\BibitemShut {NoStop}%
\bibitem [{\citenamefont {Naidon}\ and\ \citenamefont
  {Endo}(2017)}]{Naidon2017}%
  \BibitemOpen
  \bibfield  {author} {\bibinfo {author} {\bibfnamefont {P.}~\bibnamefont
  {Naidon}}\ and\ \bibinfo {author} {\bibfnamefont {S.}~\bibnamefont {Endo}},\
  }\bibfield  {title} {\bibinfo {title} {Efimov physics: a review},\ }\href
  {https://doi.org/10.1088/1361-6633/aa50e8} {\bibfield  {journal} {\bibinfo
  {journal} {Rep. Prog. Phys.}\ }\textbf {\bibinfo {volume} {80}},\ \bibinfo
  {pages} {056001} (\bibinfo {year} {2017})}\BibitemShut {NoStop}%
\bibitem [{\citenamefont {Goy}\ \emph {et~al.}(1995)\citenamefont {Goy},
  \citenamefont {Richard},\ and\ \citenamefont {Fleck}}]{Goy1995}%
  \BibitemOpen
  \bibfield  {author} {\bibinfo {author} {\bibfnamefont {J.}~\bibnamefont
  {Goy}}, \bibinfo {author} {\bibfnamefont {J.-M.}\ \bibnamefont {Richard}},\
  and\ \bibinfo {author} {\bibfnamefont {S.}~\bibnamefont {Fleck}},\ }\bibfield
   {title} {\bibinfo {title} {Weakly bound three-body systems with no bound
  subsystems},\ }\href {https://doi.org/10.1103/PhysRevA.52.3511} {\bibfield
  {journal} {\bibinfo  {journal} {Phys. Rev. A}\ }\textbf {\bibinfo {volume}
  {52}},\ \bibinfo {pages} {3511} (\bibinfo {year} {1995})}\BibitemShut
  {NoStop}%
\bibitem [{\citenamefont {Efimov}(1970)}]{Efimov1970}%
  \BibitemOpen
  \bibfield  {author} {\bibinfo {author} {\bibfnamefont {V.}~\bibnamefont
  {Efimov}},\ }\bibfield  {title} {\bibinfo {title} {Energy levels arising from
  resonant two-body forces in a three-body system},\ }\href
  {https://doi.org/https://doi.org/10.1016/0370-2693(70)90349-7} {\bibfield
  {journal} {\bibinfo  {journal} {Physics Letters B}\ }\textbf {\bibinfo
  {volume} {33}},\ \bibinfo {pages} {563} (\bibinfo {year} {1970})}\BibitemShut
  {NoStop}%
\bibitem [{\citenamefont {Efimov}(1973)}]{Efimov1973}%
  \BibitemOpen
  \bibfield  {author} {\bibinfo {author} {\bibfnamefont {V.}~\bibnamefont
  {Efimov}},\ }\bibfield  {title} {\bibinfo {title} {Energy levels of three
  resonantly interacting particles},\ }\href
  {https://doi.org/https://doi.org/10.1016/0375-9474(73)90510-1} {\bibfield
  {journal} {\bibinfo  {journal} {Nuclear Physics A}\ }\textbf {\bibinfo
  {volume} {210}},\ \bibinfo {pages} {157} (\bibinfo {year}
  {1973})}\BibitemShut {NoStop}%
\bibitem [{\citenamefont {Riisager}(2013)}]{Riisager2013}%
  \BibitemOpen
  \bibfield  {author} {\bibinfo {author} {\bibfnamefont {K.}~\bibnamefont
  {Riisager}},\ }\bibfield  {title} {\bibinfo {title} {Halos and related
  structures},\ }\href {https://doi.org/10.1088/0031-8949/2013/T152/014001}
  {\bibfield  {journal} {\bibinfo  {journal} {Physica Scripta}\ }\textbf
  {\bibinfo {volume} {2013}},\ \bibinfo {pages} {014001} (\bibinfo {year}
  {2013})}\BibitemShut {NoStop}%
\bibitem [{\citenamefont {Simon}(1976)}]{Simon1976}%
  \BibitemOpen
  \bibfield  {author} {\bibinfo {author} {\bibfnamefont {B.}~\bibnamefont
  {Simon}},\ }\bibfield  {title} {\bibinfo {title} {The bound state of weakly
  coupled schrödinger operators in one and two dimensions},\ }\href
  {https://doi.org/https://doi.org/10.1016/0003-4916(76)90038-5} {\bibfield
  {journal} {\bibinfo  {journal} {Annals of Physics}\ }\textbf {\bibinfo
  {volume} {97}},\ \bibinfo {pages} {279} (\bibinfo {year} {1976})}\BibitemShut
  {NoStop}%
\bibitem [{\citenamefont {Volosniev}\ \emph {et~al.}(2013)\citenamefont
  {Volosniev}, \citenamefont {Fedorov}, \citenamefont {Jensen},\ and\
  \citenamefont {Zinner}}]{Volosniev2013}%
  \BibitemOpen
  \bibfield  {author} {\bibinfo {author} {\bibfnamefont {A.~G.}\ \bibnamefont
  {Volosniev}}, \bibinfo {author} {\bibfnamefont {D.~V.}\ \bibnamefont
  {Fedorov}}, \bibinfo {author} {\bibfnamefont {A.~S.}\ \bibnamefont
  {Jensen}},\ and\ \bibinfo {author} {\bibfnamefont {N.~T.}\ \bibnamefont
  {Zinner}},\ }\bibfield  {title} {\bibinfo {title} {Occurrence conditions for
  two-dimensional borromean systems},\ }\href@noop {} {\bibfield  {journal}
  {\bibinfo  {journal} {The European Physical Journal D}\ }\textbf {\bibinfo
  {volume} {67}},\ \bibinfo {pages} {95} (\bibinfo {year} {2013})}\BibitemShut
  {NoStop}%
\bibitem [{\citenamefont {Volosniev}\ \emph {et~al.}(2014)\citenamefont
  {Volosniev}, \citenamefont {Fedorov}, \citenamefont {Jensen},\ and\
  \citenamefont {Zinner}}]{Volosniev2014}%
  \BibitemOpen
  \bibfield  {author} {\bibinfo {author} {\bibfnamefont {A.~G.}\ \bibnamefont
  {Volosniev}}, \bibinfo {author} {\bibfnamefont {D.~V.}\ \bibnamefont
  {Fedorov}}, \bibinfo {author} {\bibfnamefont {A.~S.}\ \bibnamefont
  {Jensen}},\ and\ \bibinfo {author} {\bibfnamefont {N.~T.}\ \bibnamefont
  {Zinner}},\ }\bibfield  {title} {\bibinfo {title} {Borromean ground state of
  fermions in two dimensions},\ }\href
  {https://doi.org/10.1088/0953-4075/47/18/185302} {\bibfield  {journal}
  {\bibinfo  {journal} {Journal of Physics B: Atomic, Molecular and Optical
  Physics}\ }\textbf {\bibinfo {volume} {47}},\ \bibinfo {pages} {185302}
  (\bibinfo {year} {2014})}\BibitemShut {NoStop}%
\bibitem [{\citenamefont {Happ}\ \emph {et~al.}(2022)\citenamefont {Happ},
  \citenamefont {Zimmermann},\ and\ \citenamefont {Efremov}}]{Happ2022}%
  \BibitemOpen
  \bibfield  {author} {\bibinfo {author} {\bibfnamefont {L.}~\bibnamefont
  {Happ}}, \bibinfo {author} {\bibfnamefont {M.}~\bibnamefont {Zimmermann}},\
  and\ \bibinfo {author} {\bibfnamefont {M.~A.}\ \bibnamefont {Efremov}},\
  }\bibfield  {title} {\bibinfo {title} {Universality of excited three-body
  bound states in one dimension},\ }\href
  {https://doi.org/10.1088/1361-6455/ac3cc8} {\bibfield  {journal} {\bibinfo
  {journal} {Journal of Physics B: Atomic, Molecular and Optical Physics}\
  }\textbf {\bibinfo {volume} {55}},\ \bibinfo {pages} {015301} (\bibinfo
  {year} {2022})}\BibitemShut {NoStop}%
\bibitem [{\citenamefont {Bloch}\ \emph {et~al.}(2008)\citenamefont {Bloch},
  \citenamefont {Dalibard},\ and\ \citenamefont {Zwerger}}]{Bloch2008}%
  \BibitemOpen
  \bibfield  {author} {\bibinfo {author} {\bibfnamefont {I.}~\bibnamefont
  {Bloch}}, \bibinfo {author} {\bibfnamefont {J.}~\bibnamefont {Dalibard}},\
  and\ \bibinfo {author} {\bibfnamefont {W.}~\bibnamefont {Zwerger}},\
  }\bibfield  {title} {\bibinfo {title} {Many-body physics with ultracold
  gases},\ }\href {https://doi.org/10.1103/RevModPhys.80.885} {\bibfield
  {journal} {\bibinfo  {journal} {Rev. Mod. Phys.}\ }\textbf {\bibinfo {volume}
  {80}},\ \bibinfo {pages} {885} (\bibinfo {year} {2008})}\BibitemShut
  {NoStop}%
\bibitem [{\citenamefont {Timmermans}\ \emph {et~al.}(1999)\citenamefont
  {Timmermans}, \citenamefont {Tommasini}, \citenamefont {Hussein},\ and\
  \citenamefont {Kerman}}]{Timmermans1999}%
  \BibitemOpen
  \bibfield  {author} {\bibinfo {author} {\bibfnamefont {E.}~\bibnamefont
  {Timmermans}}, \bibinfo {author} {\bibfnamefont {P.}~\bibnamefont
  {Tommasini}}, \bibinfo {author} {\bibfnamefont {M.}~\bibnamefont {Hussein}},\
  and\ \bibinfo {author} {\bibfnamefont {A.}~\bibnamefont {Kerman}},\
  }\bibfield  {title} {\bibinfo {title} {Feshbach resonances in atomic
  bose–einstein condensates},\ }\href
  {https://doi.org/https://doi.org/10.1016/S0370-1573(99)00025-3} {\bibfield
  {journal} {\bibinfo  {journal} {Physics Reports}\ }\textbf {\bibinfo {volume}
  {315}},\ \bibinfo {pages} {199} (\bibinfo {year} {1999})}\BibitemShut
  {NoStop}%
\bibitem [{\citenamefont {Chin}\ \emph {et~al.}(2010)\citenamefont {Chin},
  \citenamefont {Grimm}, \citenamefont {Julienne},\ and\ \citenamefont
  {Tiesinga}}]{Chin2010}%
  \BibitemOpen
  \bibfield  {author} {\bibinfo {author} {\bibfnamefont {C.}~\bibnamefont
  {Chin}}, \bibinfo {author} {\bibfnamefont {R.}~\bibnamefont {Grimm}},
  \bibinfo {author} {\bibfnamefont {P.}~\bibnamefont {Julienne}},\ and\
  \bibinfo {author} {\bibfnamefont {E.}~\bibnamefont {Tiesinga}},\ }\bibfield
  {title} {\bibinfo {title} {Feshbach resonances in ultracold gases},\ }\href
  {https://doi.org/10.1103/RevModPhys.82.1225} {\bibfield  {journal} {\bibinfo
  {journal} {Rev. Mod. Phys.}\ }\textbf {\bibinfo {volume} {82}},\ \bibinfo
  {pages} {1225} (\bibinfo {year} {2010})}\BibitemShut {NoStop}%
\bibitem [{\citenamefont {Dunjko}\ \emph {et~al.}(2011)\citenamefont {Dunjko},
  \citenamefont {Moore}, \citenamefont {Bergeman},\ and\ \citenamefont
  {Olshanii}}]{Dunjko2011}%
  \BibitemOpen
  \bibfield  {author} {\bibinfo {author} {\bibfnamefont {V.}~\bibnamefont
  {Dunjko}}, \bibinfo {author} {\bibfnamefont {M.~G.}\ \bibnamefont {Moore}},
  \bibinfo {author} {\bibfnamefont {T.}~\bibnamefont {Bergeman}},\ and\
  \bibinfo {author} {\bibfnamefont {M.}~\bibnamefont {Olshanii}},\ }\bibfield
  {title} {\bibinfo {title} {Chapter 10 - confinement-induced resonances},\
  }in\ \href
  {https://doi.org/https://doi.org/10.1016/B978-0-12-385508-4.00010-3} {\emph
  {\bibinfo {booktitle} {Advances in Atomic, Molecular, and Optical
  Physics}}},\ \bibinfo {series} {Advances In Atomic, Molecular, and Optical
  Physics}, Vol.~\bibinfo {volume} {60},\ \bibinfo {editor} {edited by\
  \bibinfo {editor} {\bibfnamefont {E.}~\bibnamefont {Arimondo}}, \bibinfo
  {editor} {\bibfnamefont {P.}~\bibnamefont {Berman}},\ and\ \bibinfo {editor}
  {\bibfnamefont {C.}~\bibnamefont {Lin}}}\ (\bibinfo  {publisher} {Academic
  Press},\ \bibinfo {year} {2011})\ pp.\ \bibinfo {pages}
  {461--510}\BibitemShut {NoStop}%
\bibitem [{\citenamefont {Greene}\ \emph {et~al.}(2017)\citenamefont {Greene},
  \citenamefont {Giannakeas},\ and\ \citenamefont
  {P\'erez-R\'{\i}os}}]{Green2017}%
  \BibitemOpen
  \bibfield  {author} {\bibinfo {author} {\bibfnamefont {C.~H.}\ \bibnamefont
  {Greene}}, \bibinfo {author} {\bibfnamefont {P.}~\bibnamefont {Giannakeas}},\
  and\ \bibinfo {author} {\bibfnamefont {J.}~\bibnamefont
  {P\'erez-R\'{\i}os}},\ }\bibfield  {title} {\bibinfo {title} {Universal
  few-body physics and cluster formation},\ }\href
  {https://doi.org/10.1103/RevModPhys.89.035006} {\bibfield  {journal}
  {\bibinfo  {journal} {Rev. Mod. Phys.}\ }\textbf {\bibinfo {volume} {89}},\
  \bibinfo {pages} {035006} (\bibinfo {year} {2017})}\BibitemShut {NoStop}%
\bibitem [{\citenamefont {Pires}\ \emph {et~al.}(2014)\citenamefont {Pires},
  \citenamefont {Ulmanis}, \citenamefont {H\"afner}, \citenamefont {Repp},
  \citenamefont {Arias}, \citenamefont {Kuhnle},\ and\ \citenamefont
  {Weidem\"uller}}]{Pires2014}%
  \BibitemOpen
  \bibfield  {author} {\bibinfo {author} {\bibfnamefont {R.}~\bibnamefont
  {Pires}}, \bibinfo {author} {\bibfnamefont {J.}~\bibnamefont {Ulmanis}},
  \bibinfo {author} {\bibfnamefont {S.}~\bibnamefont {H\"afner}}, \bibinfo
  {author} {\bibfnamefont {M.}~\bibnamefont {Repp}}, \bibinfo {author}
  {\bibfnamefont {A.}~\bibnamefont {Arias}}, \bibinfo {author} {\bibfnamefont
  {E.~D.}\ \bibnamefont {Kuhnle}},\ and\ \bibinfo {author} {\bibfnamefont
  {M.}~\bibnamefont {Weidem\"uller}},\ }\bibfield  {title} {\bibinfo {title}
  {Observation of efimov resonances in a mixture with extreme mass imbalance},\
  }\href {https://doi.org/10.1103/PhysRevLett.112.250404} {\bibfield  {journal}
  {\bibinfo  {journal} {Phys. Rev. Lett.}\ }\textbf {\bibinfo {volume} {112}},\
  \bibinfo {pages} {250404} (\bibinfo {year} {2014})}\BibitemShut {NoStop}%
\bibitem [{\citenamefont {Tung}\ \emph {et~al.}(2014)\citenamefont {Tung},
  \citenamefont {Jim\'enez-Garc\'{\i}a}, \citenamefont {Johansen},
  \citenamefont {Parker},\ and\ \citenamefont {Chin}}]{Tung2014}%
  \BibitemOpen
  \bibfield  {author} {\bibinfo {author} {\bibfnamefont {S.-K.}\ \bibnamefont
  {Tung}}, \bibinfo {author} {\bibfnamefont {K.}~\bibnamefont
  {Jim\'enez-Garc\'{\i}a}}, \bibinfo {author} {\bibfnamefont {J.}~\bibnamefont
  {Johansen}}, \bibinfo {author} {\bibfnamefont {C.~V.}\ \bibnamefont
  {Parker}},\ and\ \bibinfo {author} {\bibfnamefont {C.}~\bibnamefont {Chin}},\
  }\bibfield  {title} {\bibinfo {title} {Geometric scaling of efimov states in
  a $^{6}\mathrm{Li}\text{\ensuremath{-}}^{133}\mathrm{Cs}$ mixture},\ }\href
  {https://doi.org/10.1103/PhysRevLett.113.240402} {\bibfield  {journal}
  {\bibinfo  {journal} {Phys. Rev. Lett.}\ }\textbf {\bibinfo {volume} {113}},\
  \bibinfo {pages} {240402} (\bibinfo {year} {2014})}\BibitemShut {NoStop}%
\bibitem [{\citenamefont {Mehta}\ \emph {et~al.}(2007)\citenamefont {Mehta},
  \citenamefont {Esry},\ and\ \citenamefont {Greene}}]{Mehta_3b_1d}%
  \BibitemOpen
  \bibfield  {author} {\bibinfo {author} {\bibfnamefont {N.~P.}\ \bibnamefont
  {Mehta}}, \bibinfo {author} {\bibfnamefont {B.~D.}\ \bibnamefont {Esry}},\
  and\ \bibinfo {author} {\bibfnamefont {C.~H.}\ \bibnamefont {Greene}},\
  }\bibfield  {title} {\bibinfo {title} {Three-body recombination in one
  dimension},\ }\href {https://doi.org/10.1103/PhysRevA.76.022711} {\bibfield
  {journal} {\bibinfo  {journal} {Phys. Rev. A}\ }\textbf {\bibinfo {volume}
  {76}},\ \bibinfo {pages} {022711} (\bibinfo {year} {2007})}\BibitemShut
  {NoStop}%
\bibitem [{\citenamefont {Newton}(2014)}]{rogernewton}%
  \BibitemOpen
  \bibfield  {author} {\bibinfo {author} {\bibfnamefont {R.~G.}\ \bibnamefont
  {Newton}},\ }\href@noop {} {\emph {\bibinfo {title} {Scattering Theory of
  Waves and Particles}}},\ \bibinfo {edition} {2nd}\ ed.,\ Theoretical and
  Mathematical Physics\ (\bibinfo  {publisher} {Springer Berlin, Heidelberg},\
  \bibinfo {year} {18 April 2014})\BibitemShut {NoStop}%
\bibitem [{\citenamefont {Faddeev}(1961)}]{faddeev_scattering_1961}%
  \BibitemOpen
  \bibfield  {author} {\bibinfo {author} {\bibfnamefont {L.~D.}\ \bibnamefont
  {Faddeev}},\ }\bibfield  {title} {\bibinfo {title} {Scattering theory for a
  three-particle system},\ }\href@noop {} {\bibfield  {journal} {\bibinfo
  {journal} {Soviet Physics JETP}\ }\textbf {\bibinfo {volume} {12}},\ \bibinfo
  {pages} {1014} (\bibinfo {year} {1961})}\BibitemShut {NoStop}%
\bibitem [{\citenamefont {Sitenko}(2012)}]{sitenko}%
  \BibitemOpen
  \bibfield  {author} {\bibinfo {author} {\bibfnamefont {A.~G.}\ \bibnamefont
  {Sitenko}},\ }\href@noop {} {\emph {\bibinfo {title} {Scattering Theory}}},\
  \bibinfo {edition} {1st}\ ed.,\ Springer Series in Nuclear and Particle
  Physics\ (\bibinfo  {publisher} {Springer Berlin, Heidelberg},\ \bibinfo
  {year} {02 February 2012})\BibitemShut {NoStop}%
\bibitem [{\citenamefont {Ball}\ \emph {et~al.}(1968)\citenamefont {Ball},
  \citenamefont {Chen},\ and\ \citenamefont {Wong}}]{Ball1968}%
  \BibitemOpen
  \bibfield  {author} {\bibinfo {author} {\bibfnamefont {J.~S.}\ \bibnamefont
  {Ball}}, \bibinfo {author} {\bibfnamefont {J.~C.~Y.}\ \bibnamefont {Chen}},\
  and\ \bibinfo {author} {\bibfnamefont {D.~Y.}\ \bibnamefont {Wong}},\
  }\bibfield  {title} {\bibinfo {title} {Faddeev equations for atomic problems
  and solutions for the ($e$,h) system},\ }\href
  {https://doi.org/10.1103/PhysRev.173.202} {\bibfield  {journal} {\bibinfo
  {journal} {Phys. Rev.}\ }\textbf {\bibinfo {volume} {173}},\ \bibinfo {pages}
  {202} (\bibinfo {year} {1968})}\BibitemShut {NoStop}%
\bibitem [{\citenamefont {Happ}\ \emph {et~al.}(2019)\citenamefont {Happ},
  \citenamefont {Zimmermann}, \citenamefont {Betelu}, \citenamefont
  {Schleich},\ and\ \citenamefont {Efremov}}]{Happ2019}%
  \BibitemOpen
  \bibfield  {author} {\bibinfo {author} {\bibfnamefont {L.}~\bibnamefont
  {Happ}}, \bibinfo {author} {\bibfnamefont {M.}~\bibnamefont {Zimmermann}},
  \bibinfo {author} {\bibfnamefont {S.~I.}\ \bibnamefont {Betelu}}, \bibinfo
  {author} {\bibfnamefont {W.~P.}\ \bibnamefont {Schleich}},\ and\ \bibinfo
  {author} {\bibfnamefont {M.~A.}\ \bibnamefont {Efremov}},\ }\bibfield
  {title} {\bibinfo {title} {Universality in a one-dimensional three-body
  system},\ }\href {https://doi.org/10.1103/PhysRevA.100.012709} {\bibfield
  {journal} {\bibinfo  {journal} {Phys. Rev. A}\ }\textbf {\bibinfo {volume}
  {100}},\ \bibinfo {pages} {012709} (\bibinfo {year} {2019})}\BibitemShut
  {NoStop}%
\bibitem [{\citenamefont {Skorniakov}\ and\ \citenamefont
  {Ter-Martirosian}(1956)}]{stm}%
  \BibitemOpen
  \bibfield  {author} {\bibinfo {author} {\bibfnamefont {G.~V.}\ \bibnamefont
  {Skorniakov}}\ and\ \bibinfo {author} {\bibfnamefont {K.~A.}\ \bibnamefont
  {Ter-Martirosian}},\ }\href@noop {} {\bibfield  {journal} {\bibinfo
  {journal} {Zh. Eksp. Teor. Fiz.}\ }\textbf {\bibinfo {volume} {31}},\
  \bibinfo {pages} {775} (\bibinfo {year} {1956})}\BibitemShut {NoStop}%
\bibitem [{\citenamefont {Kartavtsev}\ \emph {et~al.}(2009)\citenamefont
  {Kartavtsev}, \citenamefont {Malykh},\ and\ \citenamefont
  {Sofianos}}]{Kartavtsev2009}%
  \BibitemOpen
  \bibfield  {author} {\bibinfo {author} {\bibfnamefont {O.~I.}\ \bibnamefont
  {Kartavtsev}}, \bibinfo {author} {\bibfnamefont {A.~V.}\ \bibnamefont
  {Malykh}},\ and\ \bibinfo {author} {\bibfnamefont {S.~A.}\ \bibnamefont
  {Sofianos}},\ }\bibfield  {title} {\bibinfo {title} {Bound states and
  scattering lengths of three two-component particles with zero-range
  interactions under one-dimensional confinement},\ }\href@noop {} {\bibfield
  {journal} {\bibinfo  {journal} {J. Exp. Theor. Phys.}\ }\textbf {\bibinfo
  {volume} {108}},\ \bibinfo {pages} {365} (\bibinfo {year}
  {2009})}\BibitemShut {NoStop}%
\bibitem [{\citenamefont {Zavin}\ and\ \citenamefont
  {Moiseyev}(2004)}]{RayaZavin_2004}%
  \BibitemOpen
  \bibfield  {author} {\bibinfo {author} {\bibfnamefont {R.}~\bibnamefont
  {Zavin}}\ and\ \bibinfo {author} {\bibfnamefont {N.}~\bibnamefont
  {Moiseyev}},\ }\bibfield  {title} {\bibinfo {title} {One-dimensional
  symmetric rectangular well: from bound to resonance via self-orthogonal
  virtual state},\ }\href {https://doi.org/10.1088/0305-4470/37/16/011}
  {\bibfield  {journal} {\bibinfo  {journal} {Journal of Physics A:
  Mathematical and General}\ }\textbf {\bibinfo {volume} {37}},\ \bibinfo
  {pages} {4619} (\bibinfo {year} {2004})}\BibitemShut {NoStop}%
\bibitem [{\citenamefont {Delves}\ and\ \citenamefont
  {Mohamed}(1985)}]{Delves_Mohamed_1985}%
  \BibitemOpen
  \bibfield  {author} {\bibinfo {author} {\bibfnamefont {L.~M.}\ \bibnamefont
  {Delves}}\ and\ \bibinfo {author} {\bibfnamefont {J.~L.}\ \bibnamefont
  {Mohamed}},\ }\href@noop {} {\emph {\bibinfo {title} {Computational Methods
  for Integral Equations}}}\ (\bibinfo  {publisher} {Cambridge University
  Press},\ \bibinfo {year} {1985})\BibitemShut {NoStop}%
\bibitem [{\citenamefont {Golub}\ and\ \citenamefont
  {Welsch}(1969)}]{Golub1969}%
  \BibitemOpen
  \bibfield  {author} {\bibinfo {author} {\bibfnamefont {G.~H.}\ \bibnamefont
  {Golub}}\ and\ \bibinfo {author} {\bibfnamefont {J.~H.}\ \bibnamefont
  {Welsch}},\ }\bibfield  {title} {\bibinfo {title} {Calculation of gauss
  quadrature rules},\ }\href
  {https://doi.org/https://doi.org/10.1090/S0025-5718-69-99647-1} {\bibfield
  {journal} {\bibinfo  {journal} {Math. Comp.}\ }\textbf {\bibinfo {volume}
  {23}},\ \bibinfo {pages} {221} (\bibinfo {year} {1969})}\BibitemShut
  {NoStop}%
\end{thebibliography}
\end{document}